\documentclass[10pt,twocolumn]{article}

\usepackage[a4paper,margin=1.75cm,columnsep=0.62cm]{geometry}
\usepackage{amsmath,amssymb,bm}
\usepackage{booktabs}
\usepackage{graphicx}
\usepackage{microtype}
\usepackage{mathtools}
\usepackage{array}
\usepackage{cite}
\usepackage[hidelinks]{hyperref}
\usepackage[nameinlink,noabbrev]{cleveref}
\usepackage{authblk}
\usepackage{caption}
\newcommand{\TempH}{T_{\mathrm H}}
\newcommand{\rh}{r_{\mathrm h}}
\newcommand{\kh}{\kappa_{\mathrm h}}
\newcommand{\chih}{\chi_{\mathrm h}}
\newcommand{\order}{\mathcal O}

\newcommand{\BBL}{\mathcal B_{\mathrm B,leg}}
\newcommand{\BBG}{\mathcal B_{\mathrm B,geom}}
\newcommand{\BBS}{\mathcal B_{\mathrm B,sum}}
\newcommand{\BF}{\mathcal B_{\mathrm F}}

\title{Prescription Dependence in Nonextensive Bose--Fermi State Counting near Black-Hole Horizons}
\author{Ke-Ming Shen\thanks{shen\_keming@ecut.edu.cn}}
\author{Ying Xiao}
\author{Yang Liu}
\affil{School of Science, East China University of Technology, Nanchang 330013, China}
\date{}

\begin{document}
\maketitle

\begin{abstract}
Nonextensive Bose and Fermi state is well studied with counting near black-hole horizons in a one-function, nonextremal, static, spherically symmetric class. 
The calculation is organized around a $q$-deformed grand ensemble whose total partition function is a $q$-product of single-level partitions. 
This identifies the bosonic occupation-number sum as the primary level partition and allows us to separate two operations that are often conflated: replacing that sum by a geometric-series form and applying the ordinary reciprocal-log identity to a $q$-logarithm. 
A finite-energy tunneling prescription yields a leading near-horizon state-counting term controlled by the surface gravity and a horizon-response factor. 
Entropy matching then gives an energy scale proportional to the Hawking temperature; the geometry factor cancels from normalized nonextensive corrections and from the Bose--Fermi ratio. 
We derive the first-order coefficients analytically and evaluate the unexpanded $q$-integrals numerically. 
The linear approximation is accurate to better than one percent when $q$ lies within roughly two percent of unity. 
All prescriptions recover the extensive limit, but the exact bosonic level sum and the geometric approximations give opposite signs for the leading deformation of the conventional $7/8$ Bose--Fermi relation. 
The result isolates a robust geometry-factorized structure and quantifies a distinct $q$-algebra systematic uncertainty.
\end{abstract}


\section{Introduction}\label{sec:intro}

Black-hole thermodynamics remains one of the most direct meeting points of gravitation, quantum field theory, and statistical mechanics. 
The Bekenstein--Hawking relation assigns an entropy $S_{\rm BH}=A/4$ to a horizon, while Hawking radiation associates a temperature with its surface gravity \cite{Bekenstein1973,Hawking1975}. 
In a tunneling description, energy conservation can be incorporated semiclassically through the replacement $M\rightarrow M-\omega$ when a quantum of energy $\omega$ is emitted \cite{ParikhWilczek2000}. 
Despite the maturity of the thermodynamic framework, the microscopic origin of black-hole entropy and the statistical description of near-horizon degrees of freedom remain open problems; a recent review is given by Mann \cite{Mann2026}.

A concrete statistical-mechanical construction is the brick-wall model of 't Hooft \cite{tHooft1985}, in which the density of quantum-field states near the horizon generates an area-like entropy. 
The original model requires a short-distance cutoff, and subsequent work has clarified the relation of this divergence to renormalization and entanglement entropy \cite{Solodukhin1995,Demers1995,Solodukhin2011,Salasnich2023}. 
This history makes two distinctions important for generalized state-counting models: a short-distance ultraviolet prescription should not be confused with a large-radius infrared regulator, and a leading near-horizon contribution should not be identified with the full entropy of the exterior thermal atmosphere.

Generalized statistics provides a second line of motivation. 
Self-gravitating systems are long-range systems and can display nonadditivity, negative heat capacities, and ensemble inequivalence \cite{Padmanabhan1990,Campa2009}. 
Tsallis statistics \cite{Tsallis1988} offers one parameterized extension of Boltzmann--Gibbs statistics, and generalized entropies have been widely explored in black-hole thermodynamics \cite{TsallisCirto2013,BiroCzinner2013,CzinnerIguchi2016,Majhi2017,Mejrhit2019,Barrow2020,Plastino2022,Zhang2024,Tong2024}. 
The area law by itself, however, does not uniquely select a generalized entropy or determine the deformation parameter. 
Recent work has therefore emphasized different microscopic or thermodynamic routes to nonextensivity. 
Maleki, Ebadi, and Mohammadzadeh studied nonextensive Bose statistics for a gas near the horizon and showed that an appropriate parameter choice can reproduce the Bekenstein--Hawking entropy \cite{Maleki2024}. 
Barzi, El Moumni, and Masmar encoded nonextensivity through a generalized Euclidean path integral and Wick rotation \cite{Barzi2025}. 
More recently, Chunaksorn, Nakarachinda, and Wongjun derived a Tsallis-modified black-hole entropy from a near-horizon classical gas and analyzed the resulting thermodynamic stability \cite{Chunaksorn2025}.

The present work addresses a complementary issue that becomes specific to quantum statistics: the internal consistency of the deformed level partition. 
Generalized Bose--Einstein and Fermi--Dirac distributions already raise nontrivial consistency questions, including particle--hole symmetry and the choice of deformed exponential \cite{BiroShenZhang2015}. 
In the generalized grand ensemble of Shen, Zhang, and Wang \cite{Shen2017}, the total partition function is constructed as a $q$-product of single-level partitions. 
For bosons, the single-level object is the occupation-number sum $Z_q(l)=\sum_{n=0}^{\infty}\exp_q(-n x_l)$. 
Their subsequent replacement of this sum by an ordinary-looking geometric expression is explicitly approximate. 
Moreover, a second ordinary identity, $\ln(x^{-1})=-\ln x$, has no undeformed analogue for the $q$-logarithm. 
These distinctions disappear at $q=1$ but not away from the extensive limit. 
They therefore provide a sharp test of which nonextensive corrections follow from the chosen ensemble and which arise from algebraic approximations.

We make four improvements over the earlier Schwarzschild-only calculation. 
First, we isolate the near-horizon structure for a one-function class of nonextremal, static, spherically symmetric metrics and show that the geometry enters through a single horizon-response factor. Second, we use the exact bosonic occupation-number sum implied by the $q$-product ensemble as the primary prescription and separate it from both the geometric-series approximation and the additional reciprocal-$q$-log substitution used in the earlier calculation. Third, we derive the first-order coefficients analytically and evaluate the unexpanded $q$ integrals numerically, thereby quantifying the range in which the linearized formulas are reliable. Fourth, we show that the geometry-dependent factor cancels from normalized nonextensive slopes and from the Bose--Fermi ratio. The resulting sign sensitivity is therefore a $q$-algebra effect rather than a peculiarity of Schwarzschild geometry.

Throughout we use natural units $c=\hbar=k_{\rm B}=G=1$. The deformation parameter is treated phenomenologically; no microscopic value of $q$ is assumed.

\section{Near-horizon state counting for a static spherical horizon}\label{sec:geometry}

\subsection{Geometry, temperature, and finite-energy backreaction}

Consider the one-function (Schwarzschild-gauge) subclass of asymptotically flat, static, spherically symmetric metrics in Einstein gravity,
\begin{equation}
 ds^2=-f(r;M,\lambda)dt^2+\frac{dr^2}{f(r;M,\lambda)}+r^2d\Omega_2^2,
 \label{eq:metric}
\end{equation}
where $\lambda$ denotes any additional fixed parameters such as charge. A nonextremal outer horizon $r=\rh(M,\lambda)$ satisfies
\begin{equation}
 f(\rh;M,\lambda)=0,\qquad
 \partial_r f\big|_{\rh}=2\kh>0,
 \label{eq:horizon}
\end{equation}
and the asymptotic Hawking temperature is $\TempH=\kh/(2\pi)$. The local Tolman inverse temperature is $\beta_{\rm loc}=\sqrt{f}\,\beta_\infty$ \cite{Tolman1930}.

For an emitted neutral quantum of energy $\omega$, we implement the same semiclassical energy-conservation step used in the tunneling picture, $M\to M-\omega$ \cite{ParikhWilczek2000}, while holding $\lambda$ fixed. Thus, when $\lambda$ includes a black-hole charge, the example below concerns a neutral emitted field. Define the horizon response
\begin{equation}
 \chih\equiv\left(\frac{\partial \rh}{\partial M}\right)_{\lambda}.
 \label{eq:chi}
\end{equation}
Differentiating the horizon condition gives $\partial_M f|_{\rh}=-2\kh\chih$. Hence, for $u=r-\rh$ and small $u,\omega$,
\begin{equation}
 f(r;M-\omega,\lambda)
 =2\kh\left(u+\chih\omega\right)
 +\order(u^2,u\omega,\omega^2).
 \label{eq:nearhorizonf}
\end{equation}
For the black-hole families considered below we assume $\chih>0$, so the new horizon lies inside the original one after a positive-energy emission. At finite $\omega$, the redshift factor therefore does not vanish at the original horizon. This observation regularizes the lower limit of the particular finite-energy state count used here, but it should not be interpreted as a general solution of the brick-wall ultraviolet divergence: the divergence reappears as $\omega\to0$. If the emitted quantum also changes a parameter in $\lambda$, the one-parameter response $\chih$ must be replaced by the corresponding multivariable horizon variation.

\subsection{Extensive state count and leading entropy}

For a free, massless field with unit spin degeneracy, the local phase-space measure gives the same radial factor for Bose and Fermi sectors. With $h=2\pi$ in natural units, the extensive Bose partition function is
\begin{equation}
 \ln Z^{(\mathrm B)}
 =\frac{2\pi^3}{45}\,T_\infty^3
 \int_{\rh}^{L}\frac{r^2\,dr}{f_\omega(r)^2},
 \label{eq:ZBext}
\end{equation}
where $f_\omega(r)=f(r;M-\omega,\lambda)$ and $L$ is a large-radius infrared regulator. At fixed geometry and $\omega$,
\begin{equation}
 S=\left(1-\beta_\infty\frac{\partial}{\partial\beta_\infty}\right)\ln Z,
 \label{eq:Sthermo}
\end{equation}
so the Bose entropy carries an overall factor four. Using \cref{eq:nearhorizonf}, the leading lower-limit contribution to the radial integral is
\begin{equation}
 \int_{\rh}^{L}\frac{r^2\,dr}{f_\omega(r)^2}
 =\frac{\rh^2}{4\kh^2\chih\omega}+\mathcal R(L,\omega),
 \label{eq:Inh}
\end{equation}
where $\mathcal R$ contains the infrared thermal atmosphere and subleading near-horizon terms. Only the displayed $1/\omega$ term is used in the entropy matching below. The leading extensive entropies are therefore
\begin{align}
 S_{\rm nh}^{(\mathrm B)}&=
 \frac{8\pi^3}{45}T_\infty^3\frac{\rh^2}{4\kh^2\chih\omega},\label{eq:SBext}\\
 S_{\rm nh}^{(\mathrm F)}&=
 \frac{7\pi^3}{45}T_\infty^3\frac{\rh^2}{4\kh^2\chih\omega}.\label{eq:SFext}
\end{align}
The usual $7/8$ factor is recovered before any nonextensive deformation is introduced.

\begin{table}[t]
\caption{Horizon response for two examples. The Reissner--Nordstr\"om derivative is taken at fixed charge, with $\Delta=\sqrt{M^2-Q^2}$. The extremal limit is excluded because \cref{eq:nearhorizonf} assumes $\kh>0$.}
\label{tab:geometry}
\centering
\small
\begin{tabular}{@{}lll@{}}
\toprule
Geometry & $\rh$ & $\chih$ \\
\midrule
Schwarzschild & $2M$ & $2$ \\
Reissner--Nordstr\"om & $M+\Delta$ & $\rh/\Delta$ \\
\bottomrule
\end{tabular}
\end{table}

\section{Nonextensive grand ensemble and the bosonic level sum}\label{sec:qensemble}

Define the $q$-logarithm and $q$-exponential by
\begin{equation}
 \ln_q x=\frac{x^{1-q}-1}{1-q},\qquad
 \exp_q x=\left[1+(1-q)x\right]_+^{1/(1-q)},
 \label{eq:qfunctions}
\end{equation}
where $[y]_+=\max(y,0)$. In the generalized grand ensemble of Ref.~\cite{Shen2017}, the total partition function is a $q$-product of single-level partitions, so that its $q$-logarithm is additive over levels. At vanishing chemical potential the level partitions are
\begin{align}
 Z_{q,l}^{(\mathrm F)}&=1+\exp_q(-x_l),\label{eq:ZFlevel}\\
 Z_{q,l}^{(\mathrm B,sum)}&=\sum_{n=0}^{\infty}\exp_q(-n x_l),\qquad x_l=\beta_{\rm loc}\epsilon_l.\label{eq:ZBsumlevel}
\end{align}
Two approximation steps must now be distinguished. The occupation sum may first be replaced by the geometric expression
\begin{equation}
 Z_{q,l}^{(\mathrm B,geom)}\equiv
 \frac{1}{1-\exp_q(-x_l)}.
 \label{eq:ZBgeom}
\end{equation}
Even after this replacement, however,
\begin{equation}
 \ln_q(x^{-1})=-x^{q-1}\ln_q x\neq-\ln_q x
 \qquad(q\neq1).
 \label{eq:qloginverse}
\end{equation}
Thus the expression $-\ln_q[1-\exp_q(-x_l)]$ used in the earlier calculation is not exactly the $q$-logarithm of \cref{eq:ZBgeom}. We therefore compare three bosonic constructions: the direct occupation sum, the geometric approximation with its $q$-logarithm taken consistently, and the earlier ``legacy factorized-log'' expression. After converting the level sum to the continuum, their dimensionless coefficients are
\begin{align}
 A_{\mathrm B,sum}(q)&=\frac{2}{\pi}\int_0^{\infty}dt\,t^2
 \ln_q\!\left[\sum_{n=0}^{\infty}\exp_q(-nt)\right],\label{eq:ABsumexact}\\
 A_{\mathrm B,geom}(q)&=\frac{2}{\pi}\int_0^{\infty}dt\,t^2
 \ln_q\!\left[\frac{1}{1-\exp_q(-t)}\right],\label{eq:ABgeomexact}\\
 A_{\mathrm B,leg}(q)&=-\frac{2}{\pi}\int_0^{\infty}dt\,t^2
 \ln_q\!\left[1-\exp_q(-t)\right],\label{eq:ABlegacyexact}\\
 A_{\mathrm F}(q)&=\frac{2}{\pi}\int_0^{\infty}dt\,t^2
 \ln_q\!\left[1+\exp_q(-t)\right].\label{eq:AFexact}
\end{align}
For $q<1$ the $q$-exponential has compact support; for $q>1$ it has a power-law tail. The large-$t$ behavior of these thermodynamic integrals requires $q<4/3$. The exact Bose level sum itself converges for $q<2$, so the thermodynamic integral is the stronger restriction in the range considered here.

With the effective inverse-temperature parameter entering the generalized ensemble identified with the local Tolman inverse temperature, define the continuum log-partition functional
\begin{equation}
 \mathcal L_q^{(a)}=A_a(q)T_\infty^3
 \int_{\rh}^{L}\frac{r^2\,dr}{f_\omega(r)^2},
 \label{eq:Lqgeneral}
\end{equation}
where $a\in\{\mathrm B,sum;\mathrm B,geom;\mathrm B,leg;\mathrm F\}$. For the direct-sum, geometric, and Fermi cases, $\mathcal L_q$ is the corresponding $q$-logarithm of the partition function. The legacy case is retained only as the effective log-partition prescription used in the earlier calculation. This convention identifies the effective inverse-temperature multiplier with the Tolman one and therefore absorbs the constant rescaling discussed in Ref.~\cite{Shen2017}; a microscopic determination of that rescaling would introduce additional model dependence. Since $A_a(q)$ is temperature independent at fixed $q$, the same thermodynamic operation gives
\begin{equation}
 S_q^{(a)}=\mathcal L_q^{(a)}-\beta_\infty\frac{\partial\mathcal L_q^{(a)}}{\partial\beta_\infty}
 =4\mathcal L_q^{(a)}.
 \label{eq:Sqgeneral}
\end{equation}

\subsection{Analytic expansion about the extensive limit}

Expanding \cref{eq:ABsumexact,eq:ABgeomexact,eq:ABlegacyexact,eq:AFexact} around $q=1$ yields
\begin{align}
 A_{\mathrm B,sum}(q)&=\frac{2\pi^3}{45}+\frac{\BBS}{\pi}(q-1)+\order((q-1)^2),\label{eq:ABsumlin}\\
 A_{\mathrm B,geom}(q)&=\frac{2\pi^3}{45}+\frac{\BBG}{\pi}(q-1)+\order((q-1)^2),\label{eq:ABgeomlin}\\
 A_{\mathrm B,leg}(q)&=\frac{2\pi^3}{45}+\frac{\BBL}{\pi}(q-1)+\order((q-1)^2),\label{eq:ABleglin}\\
 A_{\mathrm F}(q)&=\frac{7\pi^3}{180}+\frac{\BF}{\pi}(q-1)+\order((q-1)^2).\label{eq:AFlin}
\end{align}
The coefficients are
\begin{align}
 \BBS&=48\zeta(4)-32\zeta(5)\nonumber\\
 &\quad+\frac{2\pi^2}{3}\zeta(3)=26.67904445\ldots,\label{eq:BBS}\\
 \BBG&=16\zeta(5)+\frac{2\pi^2}{3}\zeta(3)\nonumber\\
 &=24.50006148\ldots,\label{eq:BBG}\\
 \BBL&=32\zeta(5)-\frac{2\pi^2}{3}\zeta(3)\nonumber\\
 &=25.27247076\ldots,\label{eq:BBL}\\
 \BF&=\frac{209}{8}\zeta(5)-\frac{\pi^2}{3}\zeta(3)\nonumber\\
 &=23.13512890\ldots.\label{eq:BF}
\end{align}
The derivations are summarized in \cref{app:coefficients}. The direct occupation-number sum is the level partition specified before either approximation in the generalized ensemble, and its coefficient therefore provides the preferred first-order Bose result within that framework.

\section{Entropy matching and geometry-independent ratios}\label{sec:matching}

For four-dimensional Einstein gravity, $S_{\rm BH}=\pi\rh^2$. Setting $T_\infty=\TempH$ and matching only the leading term of \cref{eq:Inh} gives, for any of the four statistical constructions,
\begin{equation}
 \boxed{\quad
 \omega_q^{(a)}=\frac{A_a(q)}{4\pi^3\chih}\,\TempH\quad}
 \label{eq:genericomega}
\end{equation}
within the near-horizon and finite-energy approximations stated above. The entire static-geometry dependence enters through $\chih$. For Schwarzschild, $\chih=2$, and the extensive limits are
\begin{equation}
 \omega^{(\mathrm B)}=\frac{\TempH}{180},\qquad
 \omega^{(\mathrm F)}=\frac{7\TempH}{1440}.
 \label{eq:schwext}
\end{equation}
For a Reissner--Nordstr\"om black hole at fixed $Q$,
\begin{equation}
 \omega_q^{(a)}=\frac{\Delta}{\rh}\frac{A_a(q)}{4\pi^3}\TempH,
 \qquad \Delta=\sqrt{M^2-Q^2},
 \label{eq:RNomega}
\end{equation}
which reduces smoothly to the Schwarzschild result at $Q=0$. The strict extremal limit is not covered because the linear horizon expansion requires $\kh>0$.

Equation~\eqref{eq:genericomega} makes the main universality statement transparent. The normalized deformation
\begin{equation}
 \frac{\omega_q^{(a)}}{\omega_{1}^{(a)}}
 =\frac{A_a(q)}{A_a(1)}
 \label{eq:normalizedgeneral}
\end{equation}
and the Bose--Fermi ratio
\begin{equation}
 \frac{\omega_q^{(\mathrm F)}}{\omega_q^{(\mathrm B)}}
 =\frac{A_{\mathrm F}(q)}{A_{\mathrm B}(q)}
 \label{eq:ratiogeneral}
\end{equation}
are independent of $\chih$. Hence the factorization sensitivity is not a peculiarity of Schwarzschild geometry.

At first order, the preferred direct-sum Bose prescription gives
\begin{align}
 \omega_{q,\mathrm{sum}}^{(\mathrm B)}
 &=\left[\frac{1}{90\chih}+\frac{\BBS}{4\pi^4\chih}(q-1)\right]\TempH,\label{eq:omegaBsum}\\
 \omega_q^{(\mathrm F)}
 &=\left[\frac{7}{720\chih}+\frac{\BF}{4\pi^4\chih}(q-1)\right]\TempH.\label{eq:omegaF}
\end{align}
Their difference from the extensive $7/8$ relation is
\begin{equation}
 \omega_q^{(\mathrm F)}-\tfrac{7}{8}\omega_{q,\mathrm{sum}}^{(\mathrm B)}
 =-\frac{0.05225875\ldots}{\pi^4\chih}(q-1)\TempH
 +\order((q-1)^2).
 \label{eq:preferredmismatch}
\end{equation}
For Schwarzschild, the coefficient becomes $-0.02612937\ldots/\pi^4$, reproducing the previous audit result.

If only the geometric-series approximation \cref{eq:ZBgeom} is made and its $q$-logarithm is then taken consistently, one finds
\begin{equation}
 \omega_q^{(\mathrm F)}-\tfrac{7}{8}\omega_{q,\mathrm g}^{(\mathrm B)}
 =+\frac{0.42439378\ldots}{\pi^4\chih}(q-1)\TempH
 +\order((q-1)^2).
 \label{eq:geommismatch}
\end{equation}
For the legacy factorized-log prescription, the corresponding result is
\begin{equation}
 \omega_q^{(\mathrm F)}-\tfrac{7}{8}\omega_{q,\mathrm{leg}}^{(\mathrm B)}
 =+\frac{0.25542925\ldots}{\pi^4\chih}(q-1)\TempH
 +\order((q-1)^2).
 \label{eq:legmismatch}
\end{equation}
Thus the sign reversal relative to the direct sum already appears when the geometric approximation is made; the additional reciprocal-$q$-log substitution changes its magnitude.

\section{Unexpanded \texorpdfstring{$q$}{q}-integrals and validity of the linear approximation}\label{sec:nonpert}

The analytic results above are local expansions around $q=1$. To determine their useful range, we numerically evaluate the unexpanded integrals \cref{eq:ABsumexact,eq:ABgeomexact,eq:ABlegacyexact,eq:AFexact} while keeping the same effective-temperature convention. For $q<1$, the compact support of $\exp_q$ makes the mode sums finite. For $q>1$, the direct Bose level sum is evaluated with its convergent power-law tail and the integral is restricted to $q<4/3$. No fit parameters enter this comparison.

\begin{figure*}[t]
 \centering
 \includegraphics[width=0.96\textwidth]{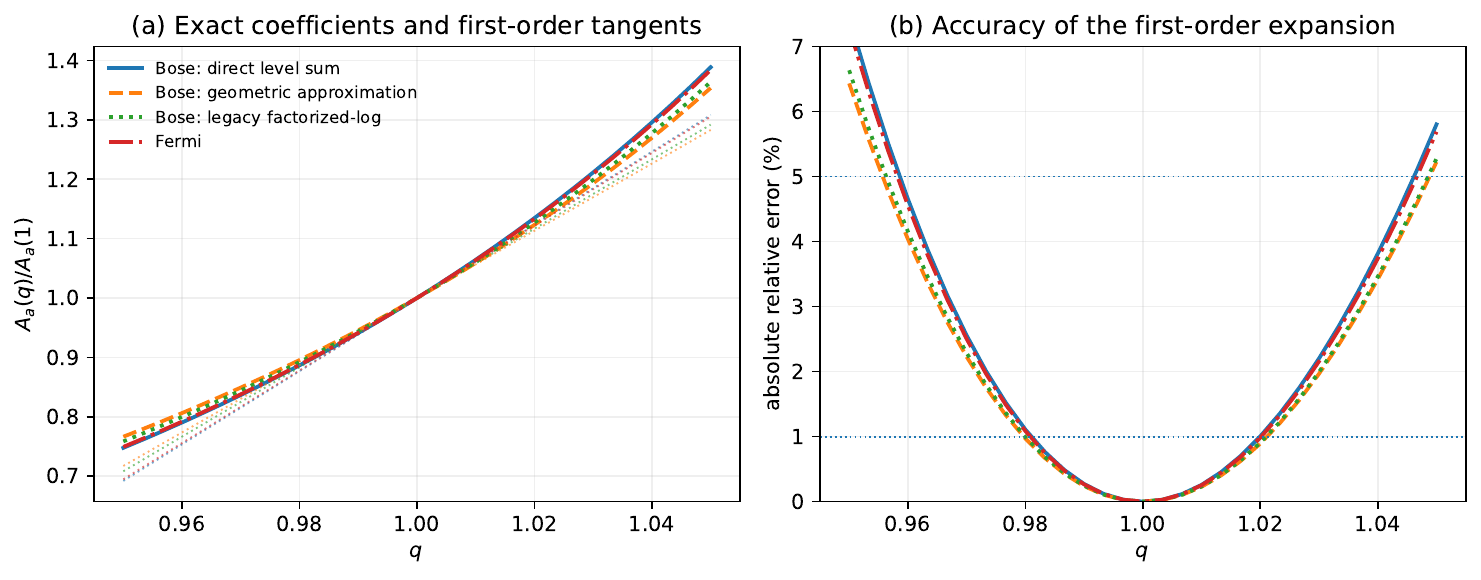}
 \caption{Nonperturbative check of the $q-1$ expansion. Panel (a) shows the unexpanded dimensionless coefficients $A_a(q)$ normalized to their extensive values. Fine dotted curves are the first-order tangents from \cref{eq:ABsumlin,eq:ABgeomlin,eq:ABleglin,eq:AFlin}. Panel (b) shows the absolute relative error of those linear approximations; the horizontal guides mark $1\%$ and $5\%$. The direct occupation-number Bose sum is the primary prescription; the geometric and legacy factorized-log curves diagnose the two approximation steps discussed in the text.}
 \label{fig:nonpert}
\end{figure*}

The numerical windows are summarized in \cref{tab:windows}. The common interval in which all three first-order approximations are accurate to better than $1\%$ is approximately
\begin{equation}
 0.981\lesssim q\lesssim1.020,
 \label{eq:onepercent}
\end{equation}
while the common $5\%$ interval is approximately $0.959\lesssim q\lesssim1.046$. This calculation gives a quantitative reason to avoid plots at $q=1.2$ when using only first-order formulas: there the truncation error is already of order twenty percent for the present coefficients.

\begin{table}[t]
\caption{Numerically determined validity windows for the first-order expansion, based on the unexpanded dimensionless integrals.}
\label{tab:windows}
\centering
\small
\begin{tabular}{@{}lcc@{}}
\toprule
Sector & $<1\%$ error & $<5\%$ error \\
\midrule
Bose, direct sum & $0.98099$--$1.01996$ & $0.95861$--$1.04618$ \\
Bose, geometric & $0.97970$--$1.02119$ & $0.95568$--$1.04885$ \\
Bose, legacy log & $0.97994$--$1.02103$ & $0.95630$--$1.04859$ \\
Fermi & $0.98079$--$1.02017$ & $0.95818$--$1.04667$ \\
\bottomrule
\end{tabular}
\end{table}

The unexpanded comparison also clarifies the physical status of the first-order sign difference. The direct, geometric, and legacy Bose constructions are tangent to different slopes at $q=1$. The direct--geometric difference is generated before the continuum integral by the non-geometric occupation sum, while the geometric--legacy difference comes from the failure of the ordinary reciprocal-log identity in $q$-algebra. Both discrepancies grow smoothly away from unity and are not numerical artifacts of the series expansion. Within the generalized $q$-product ensemble, the direct sum is therefore the preferred implementation; the other two curves quantify distinct algebraic approximations that should be treated as systematic theoretical uncertainties.

\section{Discussion}\label{sec:discussion}

Three conclusions are robust within the assumptions of the model. 
First of all, these three Bose constructions and the Fermi sector recover the ordinary quantum-statistical result at $q=1$. 
Secondly, the leading entropy-matching scale factorizes into a statistical coefficient $A_a(q)$ and a geometric horizon-response factor $1/\chih$, as shown in \cref{eq:genericomega}. 
Last, the normalized $q$ dependence and Bose--Fermi ratio are independent of that geometric factor for any nonextremal static spherical horizon in the present class.

The new element relative to recent near-horizon nonextensive studies is the algebraic sensitivity of quantum level counting. Maleki et al.~\cite{Maleki2024} focused on a nonextensive Bose gas and the recovery of horizon entropy, while Chunaksorn et al.~\cite{Chunaksorn2025} used a near-horizon classical gas to construct a Tsallis-modified black-hole entropy and study stability. The present calculation instead keeps the Bekenstein--Hawking area law as a matching condition and tests the internal $q$-algebra of the Bose/Fermi matter partition itself. In that setting, the generalized ensemble of Ref.~\cite{Shen2017} already identifies the bosonic occupation-number sum as the single-level partition and the geometric-series form as an approximation. Equations~\eqref{eq:preferredmismatch}--\eqref{eq:legmismatch} then separate the consequence of geometric factorization from the additional reciprocal-$q$-log substitution in a curved-spacetime application.

Several limitations remain important. The energy $\omega$ extracted by entropy matching is a self-consistent scale entering the backreaction-shifted state count; it is not automatically the mean or spectral-peak energy of Hawking radiation. A flux prediction would require the tunneling probability, greybody factors, spin-dependent transmission coefficients, and evolution of the background. The large-radius term $\mathcal R(L,\omega)$ has deliberately been excluded from the area-law matching; a complete exterior entropy would require a regulator-independent renormalization or entanglement-entropy treatment \cite{Demers1995,Solodukhin2011}. Finally, the effective temperature multiplier and the value of $q$ have not been derived microscopically. The generalized-ensemble parameter rescaling can itself carry $q$ dependence \cite{Shen2017}, and fixing it from a microscopic model is necessary before the absolute coefficients in \cref{eq:genericomega} can be promoted to observables. The Fermi sector here also follows that same unsymmetrized $q$-exponential ensemble at vanishing chemical potential. Particle--hole symmetry imposes additional constraints on deformed Fermi distributions \cite{BiroShenZhang2015}; incorporating a particle--hole-symmetric completion, especially for charged emission with nonzero chemical potential, is a separate extension.

\section{Conclusions}\label{sec:conclusion}

We have reformulated nonextensive near-horizon Bose--Fermi state counting in a way that separates geometry, statistical algebra, and perturbative approximation. For the one-function nonextremal static spherical class considered here, finite-energy backreaction gives a leading radial state-counting term proportional to $1/(\kh^2\chih\omega)$. After matching the associated matter entropy to the Bekenstein--Hawking area law, the emitted-energy scale takes the compact form \cref{eq:genericomega}. Geometry rescales the absolute energy through $\chih$, but cancels from normalized nonextensive corrections and from the Bose--Fermi ratio.

Within the generalized $q$-product ensemble, the exact bosonic level partition is the occupation-number sum in \cref{eq:ZBsumlevel}. Its first-order coefficient differs from both the consistently $q$-logged geometric approximation and the legacy factorized-log expression, and the induced correction to the conventional $7/8$ Bose--Fermi relation has the opposite sign. Numerical evaluation of the unexpanded $q$-integrals confirms that this difference persists beyond the local series expansion and quantifies the domain in which the first-order formulas are reliable. The principal result is therefore not a universal numerical coefficient attached to ``Tsallis black-hole entropy,'' but a cleaner statement: near-horizon nonextensive quantum statistics has a geometry-factorized structure, while the quantitative Bose correction is sensitive to both occupation-sum factorization and the reciprocal-$q$-log identity.

Future work should combine the exact level partition with greybody-resolved fluxes, determine the temperature multiplier and $q$ from a microscopic model, and test rotating or dynamical horizons where the one-parameter response $\chih$ is replaced by a multivariable first-law structure.

\section*{Acknowledgements}
The authors thank Y. J. Huang, Z. A. Jiang, Y. X. Peng, and H. H. Shi for helpful discussions.

\section*{Statements and Declarations}
\noindent\textbf{Funding.} This work did not receive any specific grant from funding agencies in the public, commercial, or not-for-profit sectors.

\noindent\textbf{Competing interests.} The authors declare no competing interests.

\noindent\textbf{Data and code availability.} No experimental or observational data were used. The numerical figure is generated from the dimensionless integrals in \cref{eq:ABsumexact,eq:ABgeomexact,eq:ABlegacyexact,eq:AFexact}; the script and tabulated values used for \cref{fig:nonpert} are included with the manuscript source package.

\appendix
\numberwithin{equation}{section}
\renewcommand{\theequation}{\thesection\arabic{equation}}
\section{First-order statistical coefficients}\label{app:coefficients}

Let $\delta=1-q$ and $t=\beta\epsilon$. The useful expansions are
\begin{align}
 \exp_q(-t)&=e^{-t}\left[1-\frac{\delta}{2}t^2+\order(\delta^2)\right],\label{eq:expqexp}\\
 \ln_q x&=\ln x+\frac{\delta}{2}(\ln x)^2+\order(\delta^2).\label{eq:lnqexp}
\end{align}
For the legacy factorized-log Bose expression, the extensive integral is $\int_0^\infty t^2[-\ln(1-e^{-t})]dt=\pi^4/45$. The $q$-exponential and $q$-logarithm corrections give, respectively,
\begin{align}
 I_{\mathrm B}^{(\exp)}&=-12\zeta(5)\,\delta,\\
 I_{\mathrm B}^{(\ln)}&=-\left[4\zeta(5)-\frac{\pi^2}{3}\zeta(3)\right]\delta.
\end{align}
Including the phase-space factor $2/\pi$ yields \cref{eq:BBL}. If the geometric approximation is instead $q$-logged consistently, the $q$-logarithm correction changes sign because of \cref{eq:qloginverse}. The net coefficient is therefore $\BBG=16\zeta(5)+(2\pi^2/3)\zeta(3)$, giving \cref{eq:BBG}.

For fermions,
\begin{align}
 \int_0^\infty t^2\ln(1+e^{-t})dt&=\frac{7\pi^4}{360},\\
 I_{\mathrm F}^{(\exp)}&=-\frac{45}{4}\zeta(5)\,\delta,\\
 I_{\mathrm F}^{(\ln)}&=\left[\frac{\pi^2}{6}\zeta(3)-\frac{29}{16}\zeta(5)\right]\delta,
\end{align}
which gives \cref{eq:BF}.

For the direct bosonic level sum, expand first
\begin{equation}
 Z_{q,l}^{\mathrm B,sum}=\sum_{n=0}^{\infty}\exp_q(-nt)
\end{equation}
and only then take the $q$-logarithm. To first order,
\begin{align}
 \ln_q Z_{q,l}^{\mathrm B,sum}
 =&-\ln(1-e^{-t})\nonumber\\
 &-\frac{1-q}{2}\,t^2
 \frac{e^{-t}(1+e^{-t})}{(1-e^{-t})^2}\nonumber\\
 &+\frac{1-q}{2}\ln^2(1-e^{-t})
 +\order((q-1)^2).
 \label{eq:directexpansion}
\end{align}
The continuum integral then gives \cref{eq:BBS}. This ordering of operations is precisely what is lost when the occupation-number sum is replaced by a geometric series before the $q$-logarithm is taken.

\section{Numerical evaluation of the unexpanded level sums}\label{app:numerics}

The unexpanded coefficients in \cref{eq:ABsumexact,eq:ABgeomexact,eq:ABlegacyexact,eq:AFexact} are one-dimensional dimensionless integrals. For $q<1$, the compact-support convention in \cref{eq:qfunctions} truncates the bosonic occupation sum at finite $n$ for every $t>0$. For $q>1$, the tail behaves as $(1+(q-1)nt)^{-1/(q-1)}$; the sum is convergent for $q<2$, and its Euler--Maclaurin tail can be evaluated analytically. The outer $t$ integral behaves as $t^{2-1/(q-1)}$ at large $t$, giving the stronger condition $q<4/3$.

Adaptive quadrature was used on the interval $0.95\le q\le1.05$. Near $t=0$, the direct bosonic level partition was stabilized with the Euler--Maclaurin expansion
\begin{equation}
 Z_{q,l}^{\mathrm B,sum}
 =\frac{1}{(2-q)t}+\frac{1}{2}+\frac{t}{12}+\order(t^3),
\end{equation}
whose contribution is multiplied by $t^2$ in \cref{eq:ABsumexact}. The numerical derivatives at $q=1$ reproduce the analytic slopes in \cref{eq:ABsumlin,eq:ABgeomlin,eq:ABleglin,eq:AFlin}. The CSV table and script included in the source package provide the values used in \cref{fig:nonpert,tab:windows}.


\begin{thebibliography}{99}
\bibitem{Bekenstein1973} J.D. Bekenstein, Black holes and entropy, Phys. Rev. D \textbf{7}, 2333 (1973). \url{https://doi.org/10.1103/PhysRevD.7.2333}.
\bibitem{Hawking1975} S.W. Hawking, Particle creation by black holes, Commun. Math. Phys. \textbf{43}, 199 (1975). \url{https://doi.org/10.1007/BF02345020}.
\bibitem{ParikhWilczek2000} M.K. Parikh, F. Wilczek, Hawking radiation as tunneling, Phys. Rev. Lett. \textbf{85}, 5042 (2000). \url{https://doi.org/10.1103/PhysRevLett.85.5042}.
\bibitem{Mann2026} R.B. Mann, Black-hole thermodynamics, Nat. Rev. Phys. \textbf{8}, 425 (2026). \url{https://doi.org/10.1038/s42254-026-00942-9}.
\bibitem{tHooft1985} G. 't Hooft, On the quantum structure of a black hole, Nucl. Phys. B \textbf{256}, 727 (1985). \url{https://doi.org/10.1016/0550-3213(85)90418-3}.
\bibitem{Solodukhin1995} S.N. Solodukhin, The conical singularity and quantum corrections to entropy of black hole, Phys. Rev. D \textbf{51}, 609 (1995). \url{https://doi.org/10.1103/PhysRevD.51.609}.
\bibitem{Demers1995} J.-G. Demers, R. Lafrance, R.C. Myers, Black hole entropy without brick walls, Phys. Rev. D \textbf{52}, 2245 (1995). \url{https://doi.org/10.1103/PhysRevD.52.2245}.
\bibitem{Solodukhin2011} S.N. Solodukhin, Entanglement entropy of black holes, Living Rev. Relativ. \textbf{14}, 8 (2011). \url{https://doi.org/10.12942/lrr-2011-8}.
\bibitem{Salasnich2023} L. Salasnich, Density of states for black holes and the brick-wall model, Symmetry \textbf{15}, 350 (2023). \url{https://doi.org/10.3390/sym15020350}.
\bibitem{Padmanabhan1990} T. Padmanabhan, Statistical mechanics of gravitating systems, Phys. Rep. \textbf{188}, 285 (1990). \url{https://doi.org/10.1016/0370-1573(90)90051-3}.
\bibitem{Campa2009} A. Campa, T. Dauxois, S. Ruffo, Statistical mechanics and dynamics of solvable models with long-range interactions, Phys. Rep. \textbf{480}, 57 (2009). \url{https://doi.org/10.1016/j.physrep.2009.07.001}.
\bibitem{Tsallis1988} C. Tsallis, Possible generalization of Boltzmann--Gibbs statistics, J. Stat. Phys. \textbf{52}, 479 (1988). \url{https://doi.org/10.1007/BF01016429}.
\bibitem{TsallisCirto2013} C. Tsallis, L.J.L. Cirto, Black hole thermodynamical entropy, Eur. Phys. J. C \textbf{73}, 2487 (2013). \url{https://doi.org/10.1140/epjc/s10052-013-2487-6}.
\bibitem{BiroCzinner2013} T.S. Bir\'o, V.G. Czinner, A $q$-parameter bound for particle spectra based on black hole thermodynamics with R\'enyi entropy, Phys. Lett. B \textbf{726}, 861 (2013). \url{https://doi.org/10.1016/j.physletb.2013.09.032}.
\bibitem{CzinnerIguchi2016} V.G. Czinner, H. Iguchi, Thermodynamics, stability and Hawking--Page transition of Kerr black holes from R\'enyi statistics, Phys. Lett. B \textbf{752}, 306 (2016). \url{https://doi.org/10.1016/j.physletb.2015.11.061}.
\bibitem{Majhi2017} B.R. Majhi, Non-extensive statistical mechanics and black hole entropy from quantum geometry, Phys. Lett. B \textbf{775}, 32 (2017). \url{https://doi.org/10.1016/j.physletb.2017.10.043}.
\bibitem{Mejrhit2019} K. Mejrhit, S.-E. Ennadifi, Thermostatistics of Schwarzschild black hole and non-Gaussian entropies, Phys. Lett. B \textbf{794}, 45 (2019). \url{https://doi.org/10.1016/j.physletb.2019.03.055}.
\bibitem{Barrow2020} J.D. Barrow, The area of a rough black hole, Phys. Lett. B \textbf{808}, 135643 (2020). \url{https://doi.org/10.1016/j.physletb.2020.135643}.
\bibitem{Plastino2022} A. Plastino, D. Monteoliva, M.C. Rocca, Black-hole thermodynamics and R\'enyi entropy, Physica A \textbf{589}, 126597 (2022). \url{https://doi.org/10.1016/j.physa.2021.126597}.
\bibitem{Zhang2024} M.Y. Zhang, H. Chen, H. Hassanabadi, Z.W. Long, H. Yang, Thermodynamic topology of Schwarzschild black holes in nonextensive statistics, Phys. Lett. B \textbf{856}, 138885 (2024). \url{https://doi.org/10.1016/j.physletb.2024.138885}.
\bibitem{Tong2024} C.W. Tong, B.H. Wang, J.R. Sun, Topology of black hole thermodynamics via R\'enyi statistics, Eur. Phys. J. C \textbf{84}, 826 (2024). \url{https://doi.org/10.1140/epjc/s10052-024-13170-1}.
\bibitem{Maleki2024} M. Maleki, Z. Ebadi, H. Mohammadzadeh, Nonextensive statistics and the entropy on the horizon, Int. J. Mod. Phys. A \textbf{39}, 2450019 (2024). \url{https://doi.org/10.1142/S0217751X24500192}.
\bibitem{Barzi2025} F. Barzi, H. El Moumni, K. Masmar, Nonextensive black hole thermodynamics from generalized Euclidean path integral and Wick's rotation, Eur. Phys. J. C \textbf{85}, 61 (2025). \url{https://doi.org/10.1140/epjc/s10052-025-13816-8}.
\bibitem{Chunaksorn2025} P. Chunaksorn, R. Nakarachinda, P. Wongjun, Black hole thermodynamics via Tsallis statistical mechanics, Eur. Phys. J. C \textbf{85}, 532 (2025). \url{https://doi.org/10.1140/epjc/s10052-025-14239-1}.
\bibitem{BiroShenZhang2015} T.S. Bir\'o, K.-M. Shen, B.-W. Zhang, Non-extensive quantum statistics with particle--hole symmetry, Physica A \textbf{428}, 410 (2015). \url{https://doi.org/10.1016/j.physa.2015.01.072}.
\bibitem{Shen2017} K.-M. Shen, B.-W. Zhang, E.-K. Wang, Generalized ensemble theory with non-extensive statistics, Physica A \textbf{487}, 215 (2017). \url{https://doi.org/10.1016/j.physa.2017.06.030}.
\bibitem{Tolman1930} R.C. Tolman, P. Ehrenfest, Temperature equilibrium in a static gravitational field, Phys. Rev. \textbf{36}, 1791 (1930). \url{https://doi.org/10.1103/PhysRev.36.1791}.
\end{thebibliography}
\end{document}